\documentclass[a4paper,12pt]{article}
\pdfoutput=1
\usepackage[margin=2.6cm]{geometry}
\usepackage{graphicx}
\usepackage[english]{babel}
\usepackage{amsmath}
\usepackage{natbib}
\usepackage[unicode=true, colorlinks=true, linkcolor=blue, citecolor=blue, urlcolor=blue, hyperfootnotes=false]{hyperref}
\usepackage{authblk}
\usepackage{tikz}
\usetikzlibrary{decorations.pathreplacing}
\bibpunct{[}{]}{,}{n}{,}{,}

\begin{document}

\title{Detection and localization of change points in temporal networks with the aid of stochastic block models}
\author[1]{Simon De Ridder\thanks{simon.deridder@ugent.be}}
\author[2,3]{Benjamin Vandermarliere\thanks{benjamin.vandermarliere@ugent.be}}
\author[2]{Jan Ryckebusch\thanks{jan.ryckebusch@ugent.be}}
\affil[1]{Department of Information Technology, Ghent University}
\affil[2]{Department of Physics and Astronomy, Ghent University}
\affil[3]{Department of General Economics, Ghent University}

\date{\today}

\maketitle

\begin{abstract}
\noindent
A framework based on generalized hierarchical random graphs (GHRGs) for the detection of change points in the structure of temporal networks has recently been developed by Peel and Clauset~\cite{Peel15}. We build on this methodology and extend it to also include the versatile stochastic block models (SBMs) as a parametric family for reconstructing the empirical networks. We use five different techniques for change point detection  on prototypical temporal networks, including empirical and synthetic ones. We find that none of the considered methods can consistently outperform the others when it comes to detecting and locating the expected change points in empirical temporal networks. With respect to the precision and the recall of the results of the change points, we find that the method based on a degree-corrected SBM has better recall properties than other dedicated methods, especially for sparse networks and smaller sliding time window widths.
\end{abstract}



\section{Introduction}

Networks are currently widely used to map and study interacting systems of animate and inanimate objects \cite{newman2003structure,newman2006structure,newman2010networks}. 
Often, the methodologies and measures developed within the context of network theories allow one to identify the central players \cite{bonacich1987power,borgatti2005centrality,borgatti2006graph} and to find structures in the nodal interactions of the network \cite{moriano2013formation}.
Thereby one often identifies groups of nodes - communities - which interact more within a group than across groups \cite{fortunato2010community,blondel2008fast,chen2014overlapping}.
Other frequently obtained topologies of social networks include the core-periphery structure \cite{holme2005core,rombach2014core} with a small group of highly interconnected core nodes and a large group of peripheral nodes that do mostly interact with core nodes.
\\

As the dynamical origins of the interactions evolve over time, the topology of the network can change \cite{holme2012temporal,holme2015modern}.
For example, a social network of high-school students changes between ``normal classes'' mode and ``summer break'' mode, not to speak about what happens to the network after graduation \cite{Peixoto15}.
There are many time evolving networks, however, for which the identification of the changes in the topological structure of the network is not that obvious.
Recently, Peel and Clauset \cite{Peel15} proposed a framework to locate the structural breaks in the large-scale structure of time-evolving networks. The proposed change point detection methodology of \cite{Peel15} develops in four steps:
\begin{enumerate}
\item[(1)] Select the generalized hierarchical random graph (GHRG) parametric family of probability distributions appropriate for reconstruction of the empirical network data.
\item[(2)] Select an appropriate width $w$ of a sliding time window. 
\item[(3)] For each time window, use  the proposed parametric family of probability distributions to infer two versions for the model: one corresponding with a change of parameters at a particular instance of time within the window, and an alternate one corresponding with the null hypothesis of no change point over the entire time window.
\item[(4)] Conduct a statistical hypothesis test to determine whether the ``change'' or ``no-change'' mode provides the better fit to the empirical network data.
\end{enumerate}
In this paper we build on this methodology, but introduce also stochastic block models (SBM) as a parametric family for reconstructing the empirical network in step (1) of the above-mentioned procedure.
The SBMs have the advantage of being very flexible.
Indeed, they can capture for example both assortative and disassortative behaviour, and core-periphery networks
\cite{holland1983stochastic,peixoto2013parsimonious,Karrer11}.
An alternate method for change point detection with an adaptive time window  based on Markov chain Monte Carlo and SBMs has recently been outlined in \cite{Peixoto15}.
\\

In what follows, we first introduce the concept of SBMs to capture a given empirical network.
Next, we detail a new method to fit a model to a given empirical network and to find the change points in a sliding time window of size $w$.
In section~\ref{results} we apply our proposed methodology to a number of prototypical temporal networks. We introduce several strategies to detect change points and compare the quality of their results. First, we conduct a study with synthetic temporal networks. Next, we apply the change-point detection methods to three empirical temporal social networks: the Enron e-mail network, the MIT proximity network, and the international trade network after $1870$. For these three networks the empirical change points are documented and we compare those with the numerical predictions.


\section{Fitting stochastic block models to a network}
\label{SBM_fitting}

In its simplest form, an SBM distributes the $N$ nodes of a network into $K$ groups.
With $n_r$ we denote the prior probability that a node is classified in group $r$. Obviously, one has that $\sum_r{n_r}=1$.
Let $\mathcal{Q}_{rs}$ be the probability that a link exists between a node $u$ in block $r$ and a node $v$ in block $s$.
The parameters $\mathcal{Q}_{rs}$  form a $K\times K$ matrix $ \left( 1 \le K \le N \right)$.
We call $g_u =r$ ($g_v =s$) the block assigned to node $u$ ($v$).
With these conventions, the probability of having a link between nodes $u$ and $v$ is Bernoulli distributed with parameter $\mathcal{Q}_{g_ug_v}$.
One can determine the likelihood of a given network (as fully determined by its adjacency matrix $A$) with a given node partitioning $\{g_u \}$  given the SBM model parameters $\{n_r\}$ and $\{\mathcal{Q}_{rs}\}$.
This can be expressed either in terms of a product over all nodes, or in terms of a product over all blocks.
\begin{align}
P(A,\{g_u\}|\{n_r\},\{\mathcal{Q}_{rs}\})
&=\prod_u{n_{g_u}}\prod_{u<v}{\mathcal{Q}_{g_ug_v}^{A_{uv}}\left(1-\mathcal{Q}_{g_ug_v}\right)^{1-A_{uv}}}\nonumber\\
&=\prod_r{n_r^{N_r}}\prod_{r\le s}{\mathcal{Q}_{rs}^{m_{rs}}\left(1-\mathcal{Q}_{rs}\right)^{N_{rs}-m_{rs}}}.
\label{bernoulli_likelihood}
\end{align}
Here, $m_{rs}$ is the number of actual links between nodes in block $r$ and nodes in block $s$.
Further, $N_{rs}$ is the total number of possible links between the nodes in block $r$ and the nodes in block $s$. For multigraphs, where $A_{uv}$ can be larger than one, the distributions in the right-hand-sides of (\ref{bernoulli_likelihood}) can be replaced by Poisson distributions.
One finds for the multigraph versions of the likelihood of (\ref{bernoulli_likelihood})   
\begin{align}
P^{\text{(Poisson)}} (A, \{g_u\}|\{n_r\},\{\mathcal{Q}_{rs}\})
&=\prod_u{n_{g_u}}\prod_{u<v}{\frac{\mathcal{Q}_{g_ug_v}^{A_{uv}}e^{-\mathcal{Q}_{g_ug_v}}}{A_{uv}!}}\nonumber\\
&= \prod_r{n_r^{N_r}}\prod_{r\le s}{\mathcal{Q}_{rs}^{m_{rs}}e^{-N_{rs}\mathcal{Q}_{rs}}}\prod_{u<v}{\frac{1}{A_{uv}!}}.
\label{poisson_likelihood}
\end{align}
These expressions for the probability distributions make the SBM a powerful and versatile tool for the analysis of complex networks.
\\

With the eye on community detection in networks, one often uses the degree-corrected (DC) version of SBM~\cite{Karrer11}.
Thereby, one introduces for all nodes $u$ an extra parameter $\theta_u$ proportional to the ratio of $u$'s degree to the sum of all degrees in block $g_u$.
By doing so, the link probability $\mathcal{Q}_{g_ug_v}$ can be replaced by $\mathcal{Q}_{g_ug_v}\theta_u\theta_v$ as the probability for a link between nodes $u$ and $v$.
This replacement diminishes the dependence of  $\mathcal{Q}_{g_ug_v}$ on the magnitude of the degrees of nodes $u$ and $v$.
As a consequence, the likelihood that a node with low degree and a node with high degree belong to the same group increases, provided that their $\theta$ is low and high, respectively.
The sketched degree correction makes sure that a separation into modules is more likely than a separation into groups with similar degrees as often happens with the regular SBM version.
We refer to \cite{Karrer11} for more details concerning the degree correction.
\\

We now detail our proposed method to fit a parametric distribution to a given empirical network.
As in \cite{Decelle11,Yan14} we use belief propagation to fit an SBM to a given network.
Thereby, each node $u$ sends a ``message'' $\psi_r^{u\rightarrow v}$ to every other node $v$ in the network.
The $\psi_r^{u\rightarrow v}$ indicates the probability that node $u$ would belong to block $r$, in the absence of node $v$.
These conditional probabilities can be iteratively updated with the aid of the expression
\begin{equation}
\psi^{u\rightarrow v}_r = \frac{1}{Z^{u\rightarrow v}}n_r\prod_{w\ne u,v}{\left(\sum_s{P(A_{wu}|\mathcal{Q}_{sr})\psi^{w\rightarrow u}_s}\right)} ,
\label{message_update}
\end{equation}
with the normalization coefficient,
\begin{equation}
Z^{u\rightarrow v} =\sum_{rs}{P(A_{uv}|\mathcal{Q}_{rs})\psi^{u\rightarrow v}_r\psi^{v\rightarrow u}_s}.
\end{equation}
The marginal probability $\psi_r^u$ that node $u$ belongs to block $r$ can then be obtained from the following expression
\begin{equation}
\psi^u_r = \frac{1}{Z^u}n_r\prod_{w\ne u}{\left(\sum_s{P(A_{wu}|\mathcal{Q}_{sr})\psi^{w\rightarrow u}_s}\right)} , 
\label{marg_message_update}
\end{equation}
with the normalization coefficient
\begin{equation}
Z^u =\sum_r{n_r\prod_{w\ne u}{\left(\sum_s{P(A_{wu}|\mathcal{Q}_{sr})\psi^{w\rightarrow u}_s}\right)}} .  
\label{cp:fit:bp:free_en:eq:norms}
\end{equation}
In order to make the algorithm scalable, it is worth remarking that up to $\mathcal{O}(\frac{1}{N})$ terms, the ``messages'' between two unconnected nodes $(u,v)$ (with $A_{uv}=0$) can be approximated by the marginal probability (see \cite{Decelle11} for details) 
\begin{equation}
\psi^{u\rightarrow v}_r \approx \psi^u_r.
\label{message_sparse_approx}
\end{equation}
With this approximation, for each node $u$ one stores and updates the $\psi^u_r$  and  the $\psi^{u\rightarrow v}_r$ for $u$'s neighbours $\{v|v\ne u,A_{uv}>0\}$.
This reduces the number of ``messages'' to be updated to $N+M$, with $M$ the total number of links in the network.
Without the  approximation (\ref{message_sparse_approx}), $N^2$ probabilities $\psi^{u\rightarrow v}_r$ need to updated and stored.
\\

The ``messages'' of (\ref{message_update}) and (\ref{marg_message_update}) allow one to put forward estimates of the SBM parameters 
\begin{align}
n_r&=\left<\frac{N_r}{N}\right>=\frac{\sum_u{\psi^u_r}}{N},&\label{parameter_n_update}\\
\mathcal{Q}_{rs}&=\left<\frac{m_{rs}}{N_{rs}}\right>&\nonumber\\
&=\begin{cases}
		\frac{1}{N^2\left(\sum_{u'}{\psi^{u'}_r}\right)\left(\sum_{v'}{\psi^{v'}_s}\right)}\sum_{u\ne v}{\frac{A_{uv}P(A_{uv}|\mathcal{Q}_{rs})\psi^{u\rightarrow v}_r\psi^{v\rightarrow u}_s}{Z^{uv}}}&(r\ne s)\\
		\frac{1}{N^2\left(\sum_{u'}{\psi^{u'}_r}\right)\left(\left(\sum_{v'}{\psi^{v'}_s}\right)-1/N\right)}\sum_{u\ne v}{\frac{A_{uv}P(A_{uv}|\mathcal{Q}_{rs})\psi^{u\rightarrow v}_r\psi^{v\rightarrow u}_s}{Z^{uv}}}&(r=s).
	\end{cases}\label{parameter_Q_update}
\end{align}
Using (\ref{message_update}) one can update the ``messages'' $\{\psi^{u\rightarrow v}_r\}$ given the current estimates of the SBM parameters $\{n_r\}$ and $\{\mathcal{Q}_{rs}\}$.
The expressions (\ref{parameter_n_update}) and (\ref{parameter_Q_update}), on the other hand, provide a way to estimate the SBM parameters, given the ``messages''.
Fitting the SBM to an empirical network can then be done as follows:
\begin{enumerate}
\item[(1)] Initialise $\{\psi^{u\rightarrow v}_r\}$ for each node $u$, and the parameters $\{n_r\}$ and $\{\mathcal{Q}_{rs}\}$ randomly.
\item[(2)] Update the SBM parameters using (\ref{parameter_n_update}) and (\ref{parameter_Q_update}).
\item[(3)] Iteratively update the ``messages'' $\{\psi^{u\rightarrow v}_r\}$ and $\{ \psi^u_r \}$, using (\ref{message_update}) and (\ref{marg_message_update}) respectively, until they converge.
\item[(4)] Repeat steps (2) and (3) until both the parameters $\left( \{n_r\}, \{\mathcal{Q}_{rs}\} \right)$ and the ``messages'' $\left( \{\psi^{u\rightarrow v}_r\}, \{ \psi^u_r \} \right)$ have converged.
\end{enumerate}
This is a variant of the Expectation-Maximisation algorithm that finds the optimal parameter values using point estimates for a given initialisation.
Because this approach can cause convergence to a local minimum, it is safer to execute this algorithm multiple times with different random initialisations, and accept the solution with the highest likelihood.
\\

By using (\ref{parameter_n_update}) and  (\ref{parameter_Q_update}) we obtain  estimates of the network's parameters of which we deem that they offer some advantages over an approach 
that assigns the nodes  to blocks 
deterministically.  
This is because a node $u$ that has a high probability to reside in block $r$ ($\psi^u_r\simeq1$), retains a small probability of residing in block $s\neq r$ ($\psi^u_s>0$). Accordingly, it contributes to the estimate of $\mathcal{Q}_{ss}$ through (\ref{parameter_Q_update}). This avoids the following problem that occurs with the deterministic assignment of the nodes to blocks. Suppose that a block $s$ has a deterministically assigned set of nodes. In situations whereby those nodes have no links in the underlying network, $\mathcal{Q}_{ss}$ is estimated as zero. By the same token, using  (\ref{bernoulli_likelihood}) or (\ref{poisson_likelihood}) the likelihood of a network with one link in block $s$ is also zero. In the approach adopted in this work, the estimate of $\mathcal{Q}_{ss}$ differs from zero which implies that the likelihood of a link within block $s$ differs from zero. Indeed, this is guaranteed through the use of (\ref{parameter_n_update}) and  (\ref{parameter_Q_update}), and the fact that $\psi^u_s>0$ for all or nearly all nodes $u$. An alternate way of circumventing the sketched problem is to introduce  Bayesian priors for the $\psi^u_s$, as was done in~\cite{Peel15}. \\

We now discuss the method used to determine the number of blocks $K$. To this end, we repeat the above fitting procedure for various choices of $K$, and select the one with the minimum description length (DL). We use the definition of the DL proposed in \cite{Peixoto14}. It consists of the sum of an entropy term $\mathcal{S}$ accounting for the amount of information in the network that is described by the model, and of a model information term $\mathcal{L}$ that quantifies the information needed to describe the model. After a deterministic assignment of the nodes to blocks using $g_u=\underset{r}{\operatorname{arg\,max}}{\psi^u_r}$, the DL $\Sigma$ can be written as:
\begin{align}
\Sigma=&\sum_{r}{\ln{\Bigg(\!\!\Bigg(\!\!\!\begin{array}{c}\binom{N_r}{2}\\m_{rr}\end{array}\!\!\!\Bigg)\!\!\Bigg)}}+\sum_{r<s}{\ln{\Bigg(\!\!\Bigg(\!\!\!\begin{array}{c}N_rN_s\\m_{rs}\end{array}\!\!\!\Bigg)\!\!\Bigg)}}\nonumber\\
&+\ln{\Bigg(\!\!\Bigg(\!\!\!\begin{array}{c}\Big(\!\!\Big(\!\!\!\begin{array}{c}K\\2\end{array}\!\!\!\Big)\!\!\Big)\\M\end{array}\!\!\!\Bigg)\!\!\Bigg)}+\ln{\Bigg(\!\!\Bigg(\!\!\!\begin{array}{c}K\\M\end{array}\!\!\!\Bigg)\!\!\Bigg)}+\ln{N!}-\sum_r{N_r} \; ,
\label{descr_length}
\end{align}
where $\big(\!\binom{N}{m}\!\big)=\binom{N+m-1}{m}$ is a combination with repetitions. For directed networks, the first line of (\ref{descr_length}) becomes $\sum_{rs}{\ln{\big(\!\binom{N_rN_s}{m_{rs}}\!\big)}}$.
For the degree-corrected model, and for more information on the MDL for SBMs, we refer to \cite{peixoto2013parsimonious} and \cite{Peixoto14}. In particular, Appendix A of \cite{Peixoto14} points out that the use of the MDL is equivalent to a Bayesian model selection of the parameter $K$.


\section{Method for detection and localization of change points}
\label{change_point_detection}

\begin{figure}[htb]
	\begin{center}
		\includegraphics[width=.25\textwidth, angle = 90]{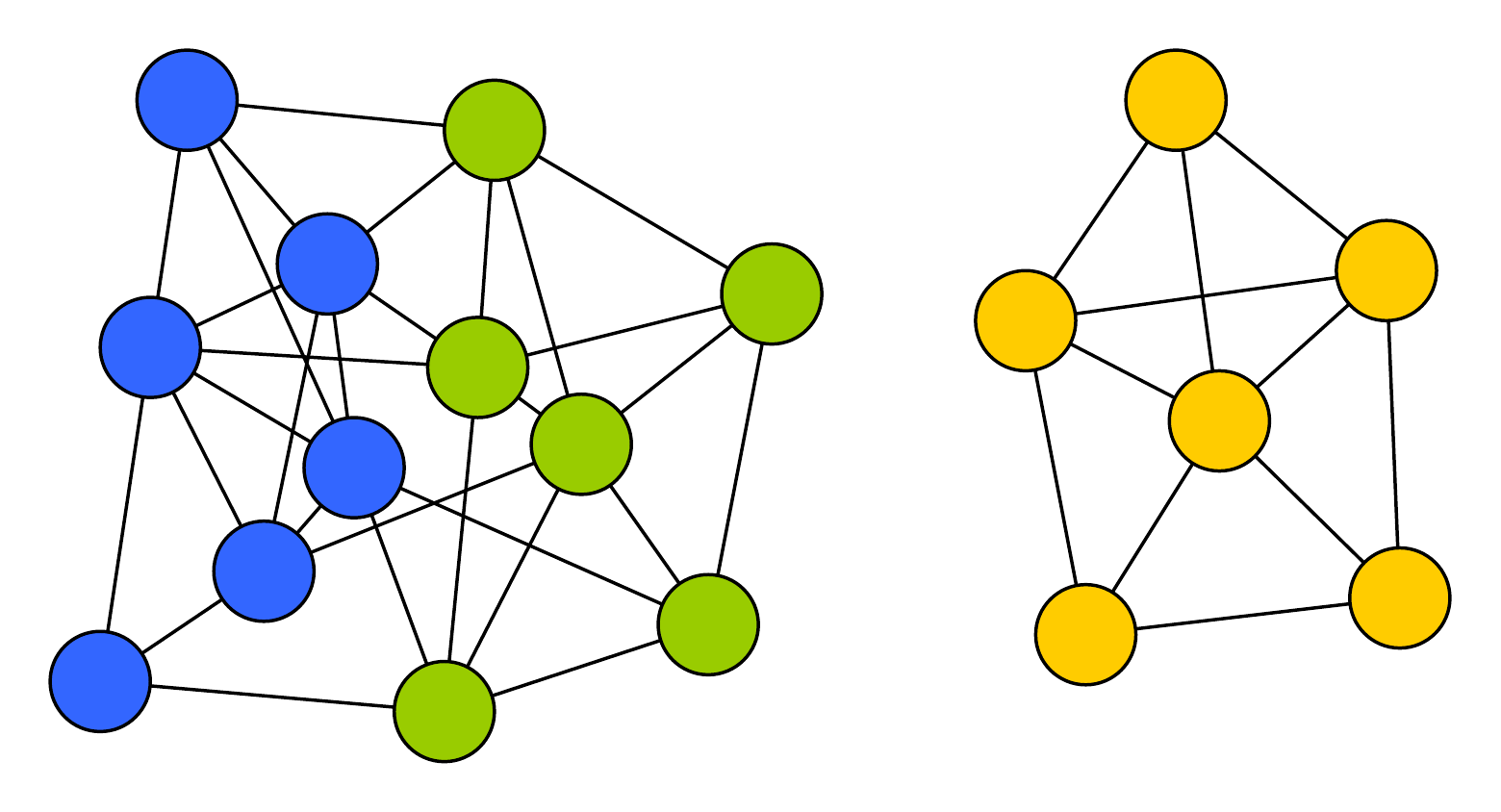}
        \includegraphics[width=.25\textwidth, angle = 90]{net_0.pdf}
   		\includegraphics[width=.25\textwidth, angle = 90]{net_0.pdf}
        \includegraphics[width=.25\textwidth, angle = 90]{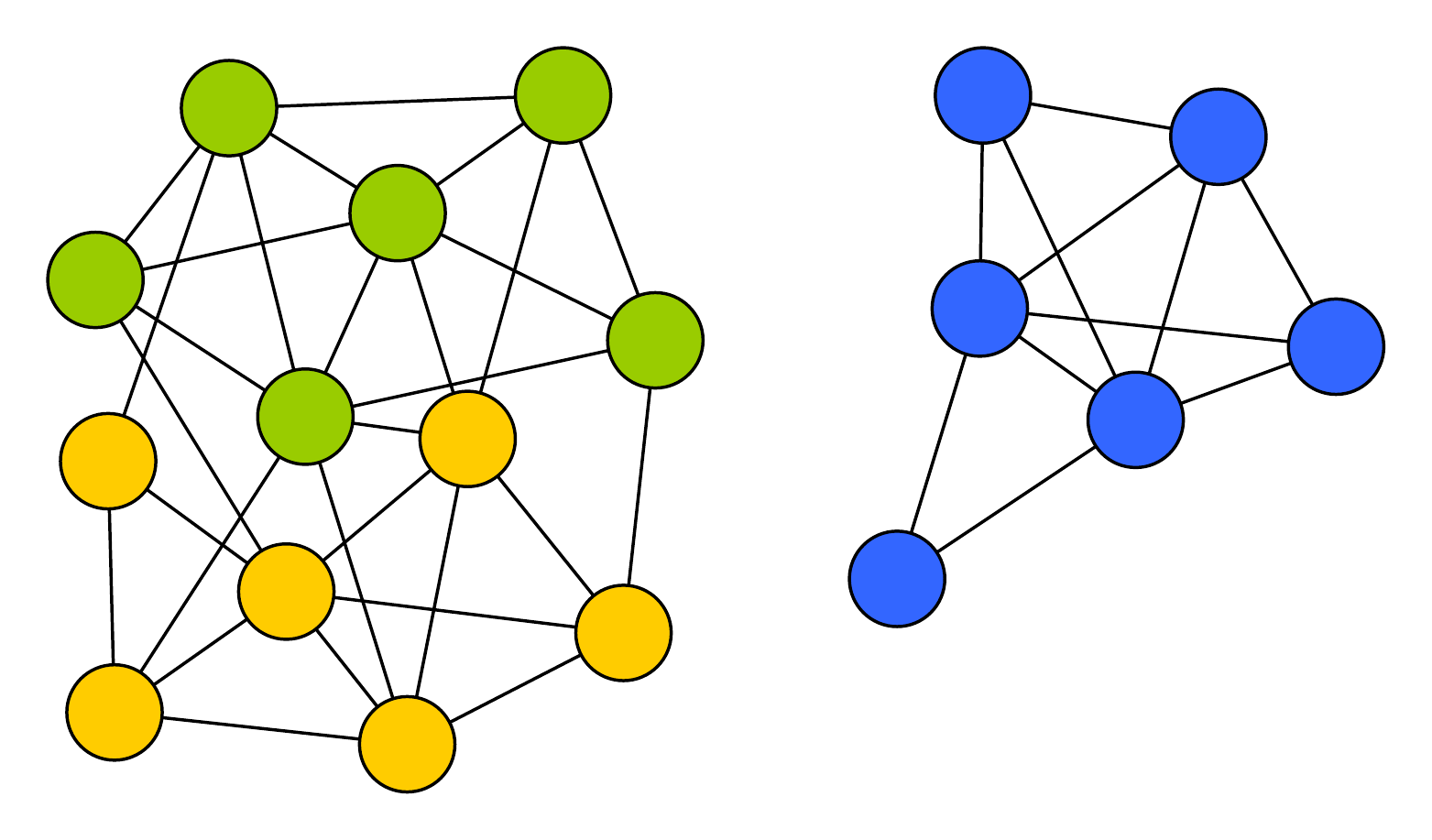}
   		\includegraphics[width=.25\textwidth, angle = 90]{net_1.pdf}\\
		\begin{tikzpicture}[node distance = 70pt]
            \node [draw=none,fill=none] (t0) {$t_0$};
            \node [draw=none,fill=none, right of=t0] (t1) {$t_1$};
            \node [draw=none,fill=none, right of=t1] (t2) {$t_2$};
            \node [draw=none,fill=none, right of=t2] (t3) {$t_3$};
            \node [draw=none,fill=none, right of=t3] (t4) {$t_4$};
            \draw [decorate,decoration={brace,amplitude=12pt,raise=4pt}]
 				  ([shift={(0.5,0)}]t2.east)--([shift={(-0.5,0)}]t0.west) node [midway,yshift=-25pt] {$\Phi_a$};
            \draw [decorate,decoration={brace,amplitude=12pt,raise=4pt}]
 				  ([shift={(0.5,0)}]t4.east)--([shift={(-0.5,0)}]t3.west)  node [midway,yshift=-25pt] {$\Phi_b$};
            \draw [decorate,decoration={brace,amplitude=12pt,raise=0pt}]
 				  ([shift={(0.5,-1)}]t4.east)--([shift={(-0.5,-1)}]t0.west)  node [midway,yshift=-20pt] {$\Phi_0$};
		\end{tikzpicture}
		\caption{Window of five consecutive snapshots ($t_0,t_1,t_2,t_3,t_4$) of a temporal network containing a change point between $t_2$ and $t_3$. Before the change point there are two distinct communities. After the change point the green nodes change sides and now make up a new community with the yellow nodes. Model $\Phi_0$ represents the null hypothesis that there is no change point in the considered time  window. A change point is detected when the combination of the two models $\Phi_a$ (fit to $(t_0, t_1, t_2)$) and $\Phi_b$ (fit to $(t_3,t_4)$) are statistically identified as a better fit to the empirical network data at five time instances.}
        \label{temporal_option2}
	\end{center}
\end{figure}

In this work, we define a temporal network as a time series of consecutive snapshots of a network.
Using the methodology of the previous section to fit an SBM to a given network, we can now proceed to develop a technique appropriate for the detection of change points in a temporal network.
The methodology rests on the idea to use an overlapping sliding time window with width $w$ and to statistically determine for each time window whether it contains a  change point or not.
With this procedure, one can detect change points without taking the full time series of networks into consideration.
\\

For each time window, we label the graphs by means of their time coordinate  $\left(t_0,t_1,\ldots,t_{w-1}\right)$ (see figure \ref{temporal_option2}).
We can test the hypothesis that a change point occurs in a particular window by considering all $w-1$ times $\left(t_1,\ldots,t_{w-1}\right)$ as possible change points.
Of those the most likely one is selected.
As a basis of reference, we start from the null hypothesis of no occurrence of a change point in the considered time window.
This hypothesis assumes no change point in the window of networks, and can therefore be based on an average model for all the networks in the window.
In order to construct such an average model in a given time window, we proceed as follows:
\begin{enumerate}
	\item[(1)] In any given time window, add all the links between every pair $(u,v)$ of nodes  and construct $A_{uv}^{[0,w-1]}=\sum_{t=t_0}^{t_{w-1}}{A_{uv}^t}$. This then forms a multigraph or a weighted network with discrete weights $0\le A_{uv}^{[0,w-1]}\le w $.
	\item[(2)] Using (\ref{poisson_likelihood}) a Poisson-distributed SBM is fitted to the obtained multigraph $A_{uv}^{[0,w-1]}$ using the belief propagation technique detailed in the previous section. Thereafter, the corresponding parameters $\{\mathcal{Q}_{rs}\}$ are divided by the window length $w$. This ensures that the expected number of links between two nodes is the average number for all network realisations in the window, rather than the sum.
\end{enumerate}
This model then forms the null model $\Phi_0$ in a conventional likelihood-ratio test.
The alternative hypothesis states that a change point occurs just before the network realisation at time instant $t_n$, with $t_0<t_n<t_w$.
For the alternative hypothesis, two other models can be constructed by re-estimating the SBM-parameters for the networks before $t_n$ (resulting in $\Phi_a$), and for the networks from $t_n$ on (resulting in $\Phi_b$) (Figure~\ref{temporal_option2}).
There are $w-1$ such hypotheses, each of which results in a log-likelihood ratio
\begin{equation}
\Lambda_{t_n}=\sum_{t=t_0}^{t_n-1}{\ln{P(A^t|\Phi_a)}}+\sum_{t=t_n}^{t_0+w-1}{\ln{P(A^t|\Phi_b)}}-\sum_{t=t_0}^{t_0+w-1}{\ln{P(A^t|\Phi_0)}}.
\label{likelihood-ratio}
\end{equation}
In order to determine the potential change point $t_n$ we select the maximum of these log-likelihood ratios,
\begin{equation}
g=\max_{t_n}{\Lambda_{t_n}} \; .
\label{g}
\end{equation}
What remains to be done is to determine whether the potential change point $t_n$ is significant.
This selection can be done by choosing a threshold value for $g$.
The traditional method to model the distribution of the log-likelihood ratios, using Wilks' theorem, is with a $\chi^2$-distribution.
It has been shown \cite{Yan14}, however, that this asymptotic approximation does not apply to a SBM.
Therefore, as in \cite{Peel15}, we make use of bootstrapping.
Bootstrapping is a way to model the distribution of the log-likelihood ratio for windows that fall under the null model, called the null distribution.
This is achieved by generating a large number of networks from the null model, and calculating the log-likelihood ratio $g'$ using (\ref{g}) for every $w$ of these networks.
As for these networks no change point should be detected, these $\{g'\}$ can be assumed to be samples from the distribution of the null model.
We can then use the distribution of these $\{g'\}$ as an approximation of the real null distribution.
A decision for the detection of a change point can then be made by selecting a confidence level and corresponding significance level, e.g.~$1-\alpha=95\!~\%$.
We calculate the $p$-value of the log-likelihood ratio $g$ as
\begin{equation}
p = \frac{\left|\{g'\}>g\right|}{\left|\{g'\}\right|} .
\end{equation}
The $p$-value determines the significance of the log-likelihood ratio, and the change point is only accepted if the condition $p<\alpha$ is met.

\section{Results}
\label{results}
In this section we present the results of our numerical studies of change-point detection. We use both synthetic (section~\ref{subsec:synthetic}) and empirical (section~\ref{subsec:empirical}) temporal networks. For all those temporal networks we use in total five methodologies to detect and locate the change points. First, the degree-corrected and the regular SBM techniques introduced in this work (DC-SBM, SBM) and the GHRG method introduced in \cite{Peel15}. We confront the results of those three involving methodologies with those of two rather straightforward local methods based on the mean degree and mean geodesic of the network. For these local methods, we calculate the specified scalars for each network in a given time window and for the network at the time instance just after the considered time window. The value for this last network is then compared to the mean value for the networks in the window, by means of a two-tailed Student's $t$-test. Thereby we adopt the same significance level $\alpha$ as used for the other methods ($1-\alpha=95\!~\%$).
\\

\subsection{Analysis with synthetic temporal networks}
\label{subsec:synthetic}

\begin{figure}[htb]
	\begin{center}
        \includegraphics[trim={7pt, 55pt, 6pt, 7pt}, clip, width=0.8\linewidth]{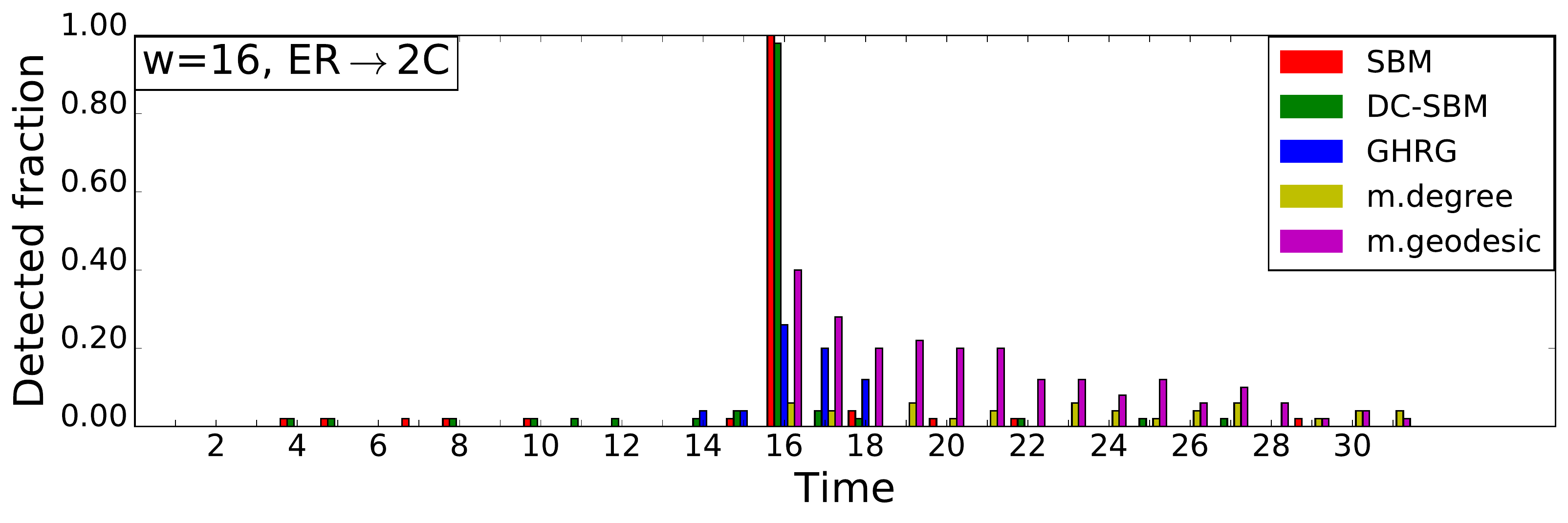}
        \includegraphics[trim={7pt, 11pt, 6pt, 7pt}, clip, width=0.8\linewidth]{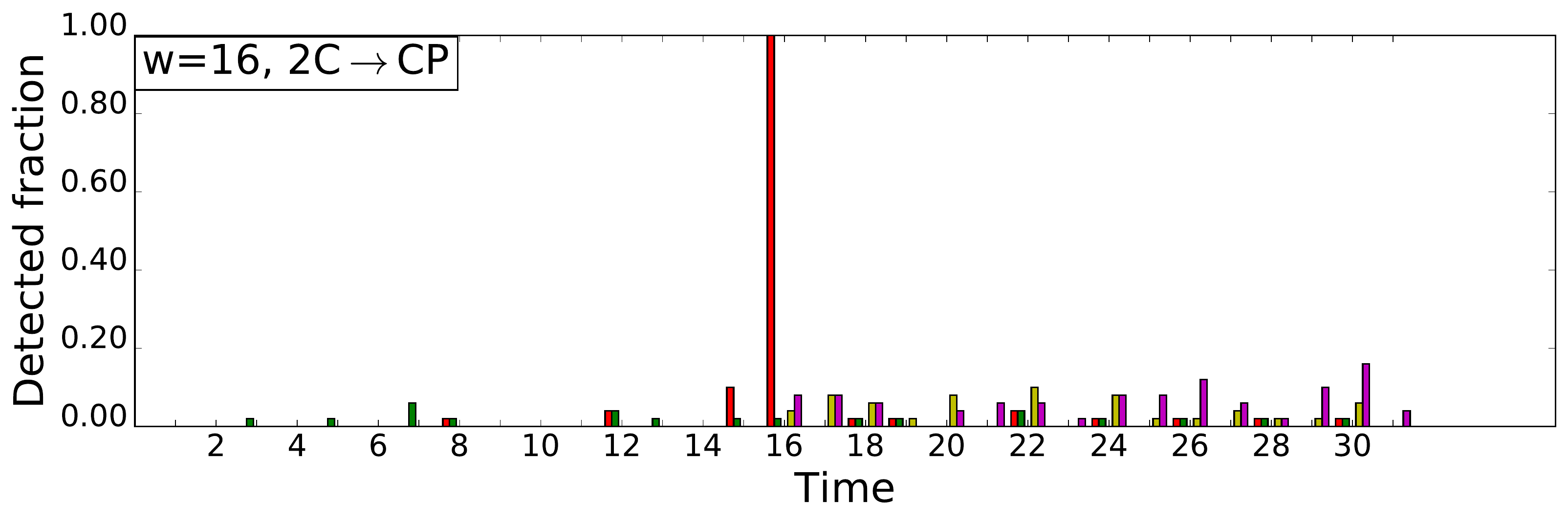} 
\caption[Synthetic CPs]{The efficiency of detecting a change point in two synthetic temporal networks with the SBM, DC-SBM, GHRG, mean-degree, and mean-geodesic methods. The true location of the change point is $t=16$. Upper panel: $t=16$ marks the change from an Erd\H{o}s-R\'enyi (ER) network to a network with two communities (2C). Lower panel: $t=16$ marks the change from a network with two communities to a network with a core-periphery (CP) structure. At all time instances, the height of the bar indicates the fraction of the 50 simulations that detect a change point. A sliding time window of size $w=16$ was used.\label{fig:synthCPs1}}
	\end{center}
\end{figure}

In this subsection we compare the performance of the proposed techniques at the retrieval of planted change points in synthetic temporal networks. We apply the methodology outlined in Sections~\ref{SBM_fitting} and \ref{change_point_detection} to the synthetic transition from an Erd\H{o}s-R\'enyi (ER) network into a network with two communities (2C), and from a network with two communities into a network with a core-periphery (CP) structure.
\\

We report results of four rounds of studies each covering $50$ simulations of $32$ time instances. Thereby, the change point is planted at $t=16$. 
The temporal synthetic networks of the ``ER'', ``2C'' and ``CP'' type are generated from their defining SBMs, with a fixed number of nodes in each block.  More specifically, the results reported are generated from:
\begin{align}
\text{ER}\rightarrow\text{2C: }&\left(\begin{array}{cc}0.1&0.1\\0.1&0.1\end{array}\right)\rightarrow\left(\begin{array}{cc}0.15&0.05\\0.05&0.15\end{array}\right), &\overline{N}=\left(\begin{array}{c}22\\28\end{array}\right),\\
\text{2C}\rightarrow\text{CP: }&\left(\begin{array}{cc}0.2&0.01\\0.01&0.2\end{array}\right)\rightarrow\left(\begin{array}{cc}0.3&0.09\\0.09&0.01\end{array}\right), &\overline{N}=\left(\begin{array}{c}20\\30\end{array}\right),\\
\text{CP}\rightarrow\text{2C: }&\left(\begin{array}{cc}0.3&0.09\\0.09&0.01\end{array}\right)\rightarrow\left(\begin{array}{cc}0.2&0.01\\0.01&0.2\end{array}\right), &\overline{N}=\left(\begin{array}{c}20\\30\end{array}\right).
\end{align}
For each simulation of a given set-up, 16 networks are independently generated from the first SBM, followed by 16 independent networks from the second SBM. This creates a time series of networks with larger variations than those typically found in the empirical temporal networks that will constitute the study of  Sec.~\ref{subsec:empirical}. We stress that the GHRG model would be an equally good choice to generate the synthetic temporal networks.\\

Figure~\ref{fig:synthCPs1} summarizes the results of the detection efficiencies for the ``ER$\rightarrow$2C'' and ``2C$\rightarrow$CP'' transitions, using a sliding window of size $w=16$, and a significance level of $1-\alpha=95\,\%$. We observe that the regular SBM method (and for the formation of two communities also the DC-SBM method) has a very high detection rate at the change point. The GHRG and the local methods have a significantly lower detection rate.
\\

\begin{figure}[htb]
	\begin{center}
        \includegraphics[trim={7pt, 55pt, 6pt, 7pt}, clip, width=0.8\linewidth]{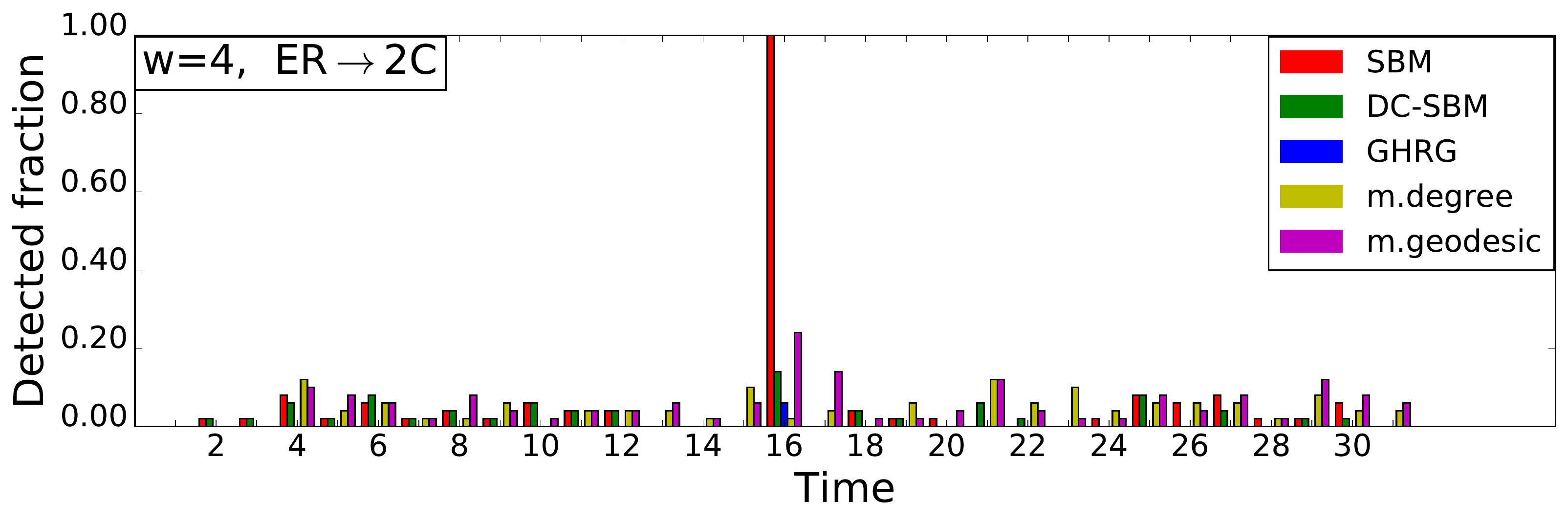}
        \includegraphics[trim={7pt, 11pt, 6pt, 7pt}, clip, width=0.8\linewidth]{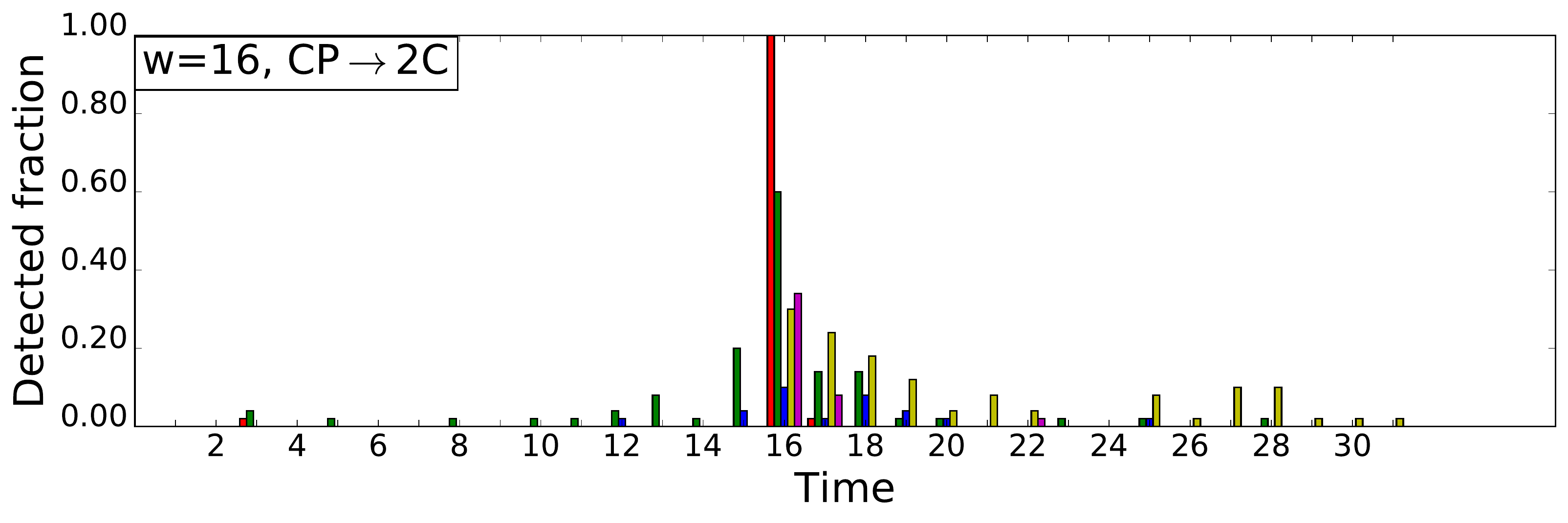} 
\caption[Synthetic CPs]{As in Fig.~\ref{fig:synthCPs1} but for a different value of $w$ (upper panel) and for the time reversed process (bottom panel). 
\label{fig:synthCPs2}}
	\end{center}
\end{figure}

Figure~\ref{fig:synthCPs2} shows the change-point detection efficiencies for two transitions related to those of Fig.~\ref{fig:synthCPs1}.  
The first is the ER$\rightarrow$2C transition with a window size $w=4$.
Comparison to the upper panel in Fig.~\ref{fig:synthCPs1} illustrates that a shorter window size causes the SBM method to predict more false predictions for local change points. We stress  that those can be partially attributed to the adopted algorithm that generates the networks independently.
Figure~\ref{fig:synthCPs2} also shows the detection efficiency results for the   CP$\rightarrow$2C transition with $w=16$.
This is the time reversed process of the one shown in the lower panel of Fig.~\ref{fig:synthCPs1}. The DC-SBM method shows a noticeable increase in detections of a change point. This is in line with the expectations, as the DC-SBM is more adept at discovering community structure than the regular SBM.  This  indicates that the DC-SBM method is better at discovering the formation of a community structure than it is at discovering its dissolution.
\\

\begin{figure}[htb]
	\begin{center}
        \includegraphics[trim={7pt, 11pt, 6pt, 7pt}, clip, width=0.9\linewidth]{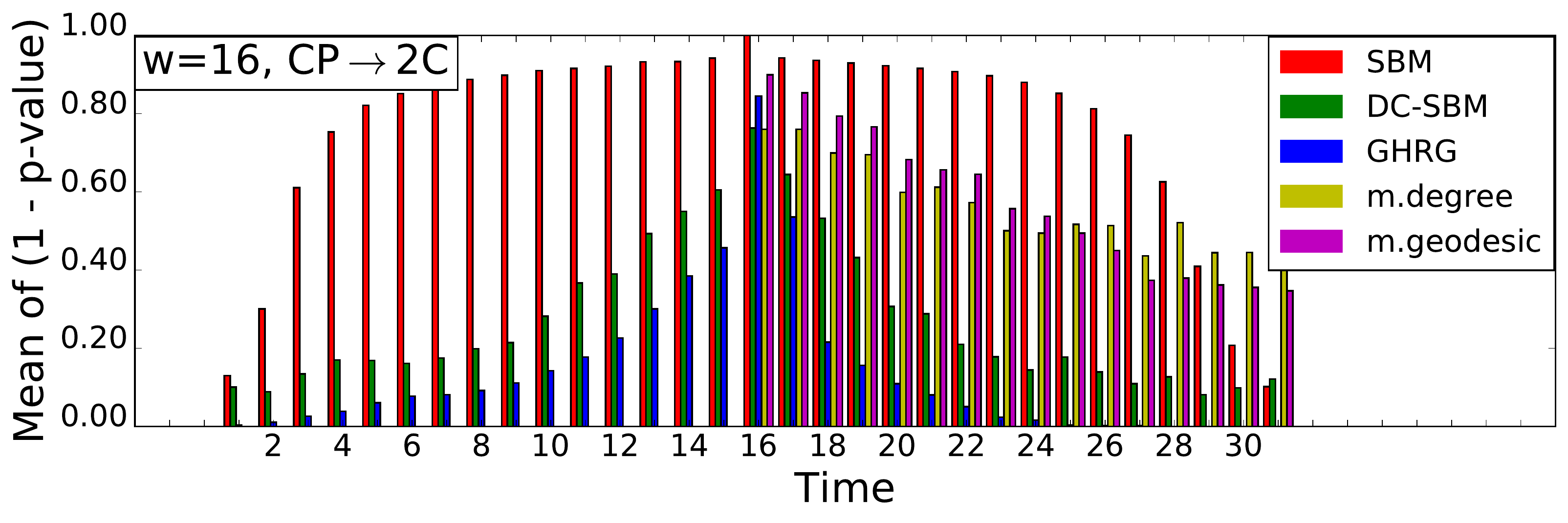}
\caption[Synthetic CPs]{The estimated probability of detecting a change point in a synthetic temporal network with the SBM, DC-SBM, GHRG, mean-degree, and mean-geodesic methods. The true location of the change point is $t=16$. It marks the change from a network with a core-periphery (CP) structure to a network with two communities. At all time instances, the height of the bar indicates one minus the p-value of the likelihood ratio statistic, averaged over all time windows containing the candidate change point for the SBM, DC-SBM and GHRG methods, and over the 50 simulations. A sliding time window of size $w=16$ was used.\label{fig:synthPVAL}}
	\end{center}
\end{figure}

In the studies summarized in Figs.~\ref{fig:synthCPs1} and \ref{fig:synthCPs2} the GHRG method seems to under-perform. The underlying reasons can be understood by inspecting Fig.~\ref{fig:synthPVAL}  showing for one specific studied transition  the mean of one minus the p-value of the likelihood ratio statistic, which can be interpreted as the probability of occurrence of a change point. We see that the GHRG, like the other methods, produces a peak in this probability, centred around the real change point. The mean, however, doesn't rise above the 95\,\% that was put forward as the   detection threshold. This indicates that at lower values of this threshold, the GHRG method would be equally efficient at predicting the $t=16$ peak. We stress that similar observations are made for all the transitions considered. 

\subsection{Analysis with empirical temporal networks}
\label{subsec:empirical}

\begin{figure}[htb]
	\begin{center}
		\includegraphics[trim={2.7cm, 3.72cm, 3.8cm, 0.8cm}, clip, width=0.65\linewidth]{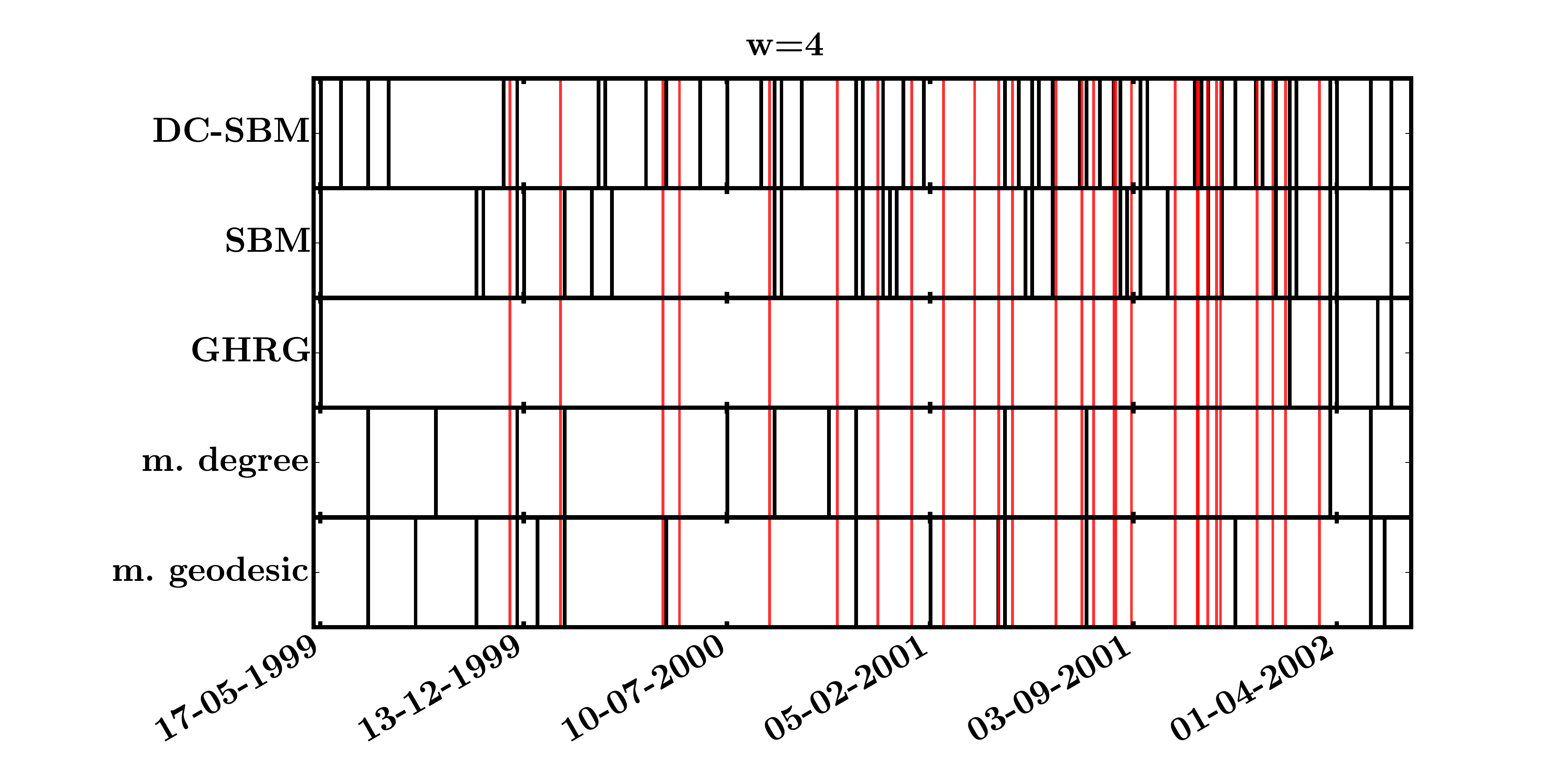}
        \includegraphics[trim={2.7cm, 1.0cm, 3.8cm, 0.3cm}, clip, width=0.65\linewidth]{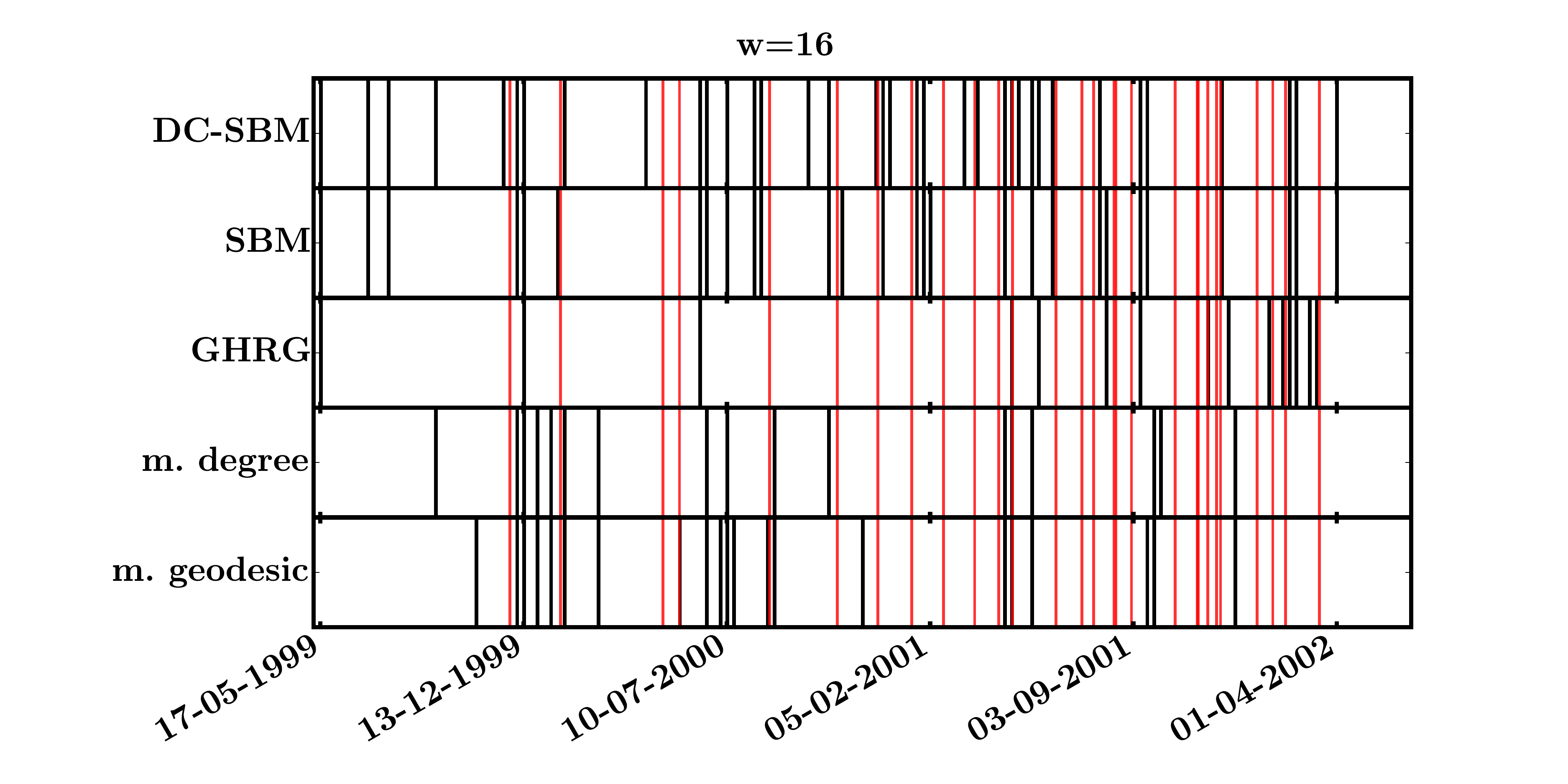}
\caption[Found change points. (Enron)]{The detected change points in the Enron e-mail network for $w=4$ weeks (upper panel) and $w=16$ weeks (lower panel). Use has been made of the SBM, DC-SBM, GHRG, mean-degree and mean-geodesic methods. The red vertical lines correspond with the time instances of documented events in the Enron company.\label{enron_cp}}
	\end{center}
\end{figure}
\begin{figure}[htb]
	\begin{center}
 		\includegraphics[trim={1.1cm, 0.6cm, 0.8cm, 0.4cm}, clip, width = 0.9\textwidth]{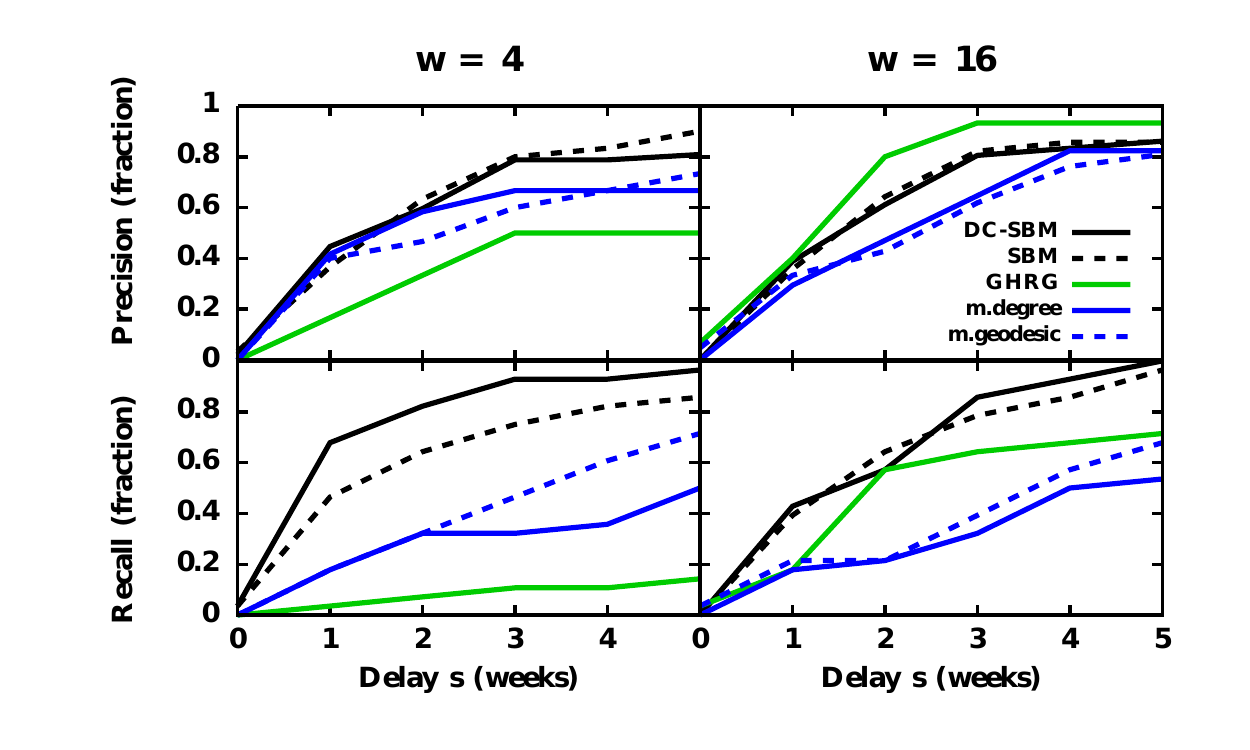}
	\end{center}
\caption[Precision and recall. (Enron)]{The computed precision (top) and recall (bottom) for the Enron e-mail network. Results are shown for window sizes of 4 (left) and 16 weeks (right) and for five change point detection methods.\label{enron_precrec}}
\end{figure}
We now apply the methodology outlined in the Sections~\ref{SBM_fitting} and \ref{change_point_detection} to three empirical temporal networks: the Enron e-mail network, the MIT proximity network and the international trade network. The first two datasets were also used in the change point analysis of~\cite{Peel15}. First, we briefly describe the three datasets that underlie the temporal networks used in our analysis.

\begin{figure}[htb]
	\begin{center}
		\includegraphics[trim={2.7cm, 3.74cm, 3.8cm, 0.8cm}, clip, width=0.65\linewidth]{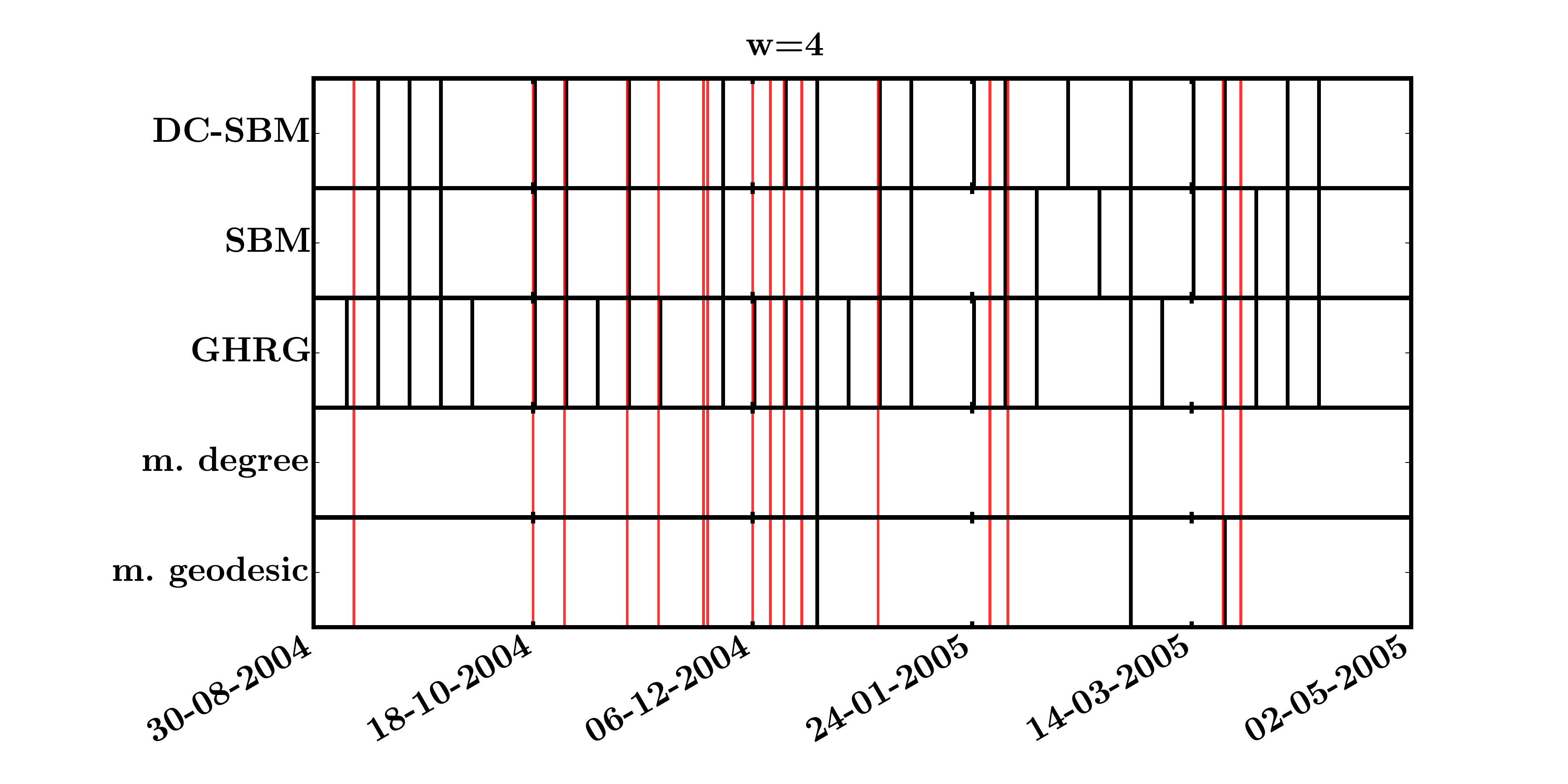}
        \includegraphics[trim={2.7cm, 1.0cm, 3.8cm, 0.3cm}, clip, width=0.65\linewidth]{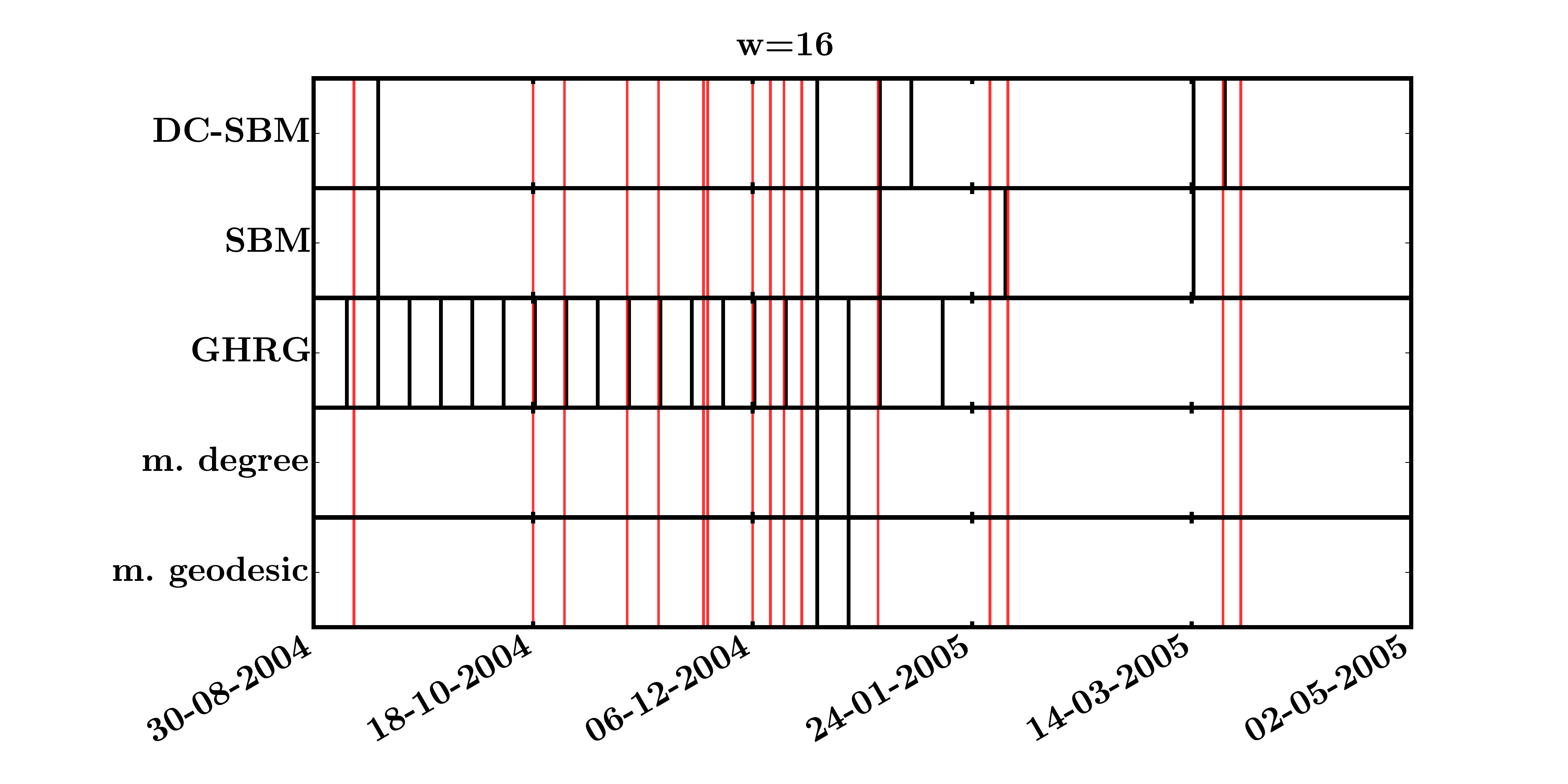}
		\caption[The change points. (MIT)]{As in Figure~\ref{enron_cp} but for the MIT proximity network.
        \label{mit_cp}}
	\end{center}
\end{figure}

Enron is a U.S.~energy company that filed for bankruptcy back in 2001 due to accounting scandals. As a result of an official inquiry, a dataset of e-mails exchanged between members of the Enron staff was made public\footnote{Available at \url{www.cs.cmu.edu/~enron}.}. With those data, one can construct a temporal network with Enron's staff members as nodes, and links which reflect the e-mail exchanges in a particular working week. In this way, one creates a sparse network with an average of 0.43 links per node.\\
The MIT reality mining project is an experiment conducted by the Media Laboratory at the Massachusetts Institute of Technology (MIT) during the 2004-2005 academic year \cite{Eagle09}. In this experiment, ninety-four subjects, both MIT students and staff, were monitored by means of their smartphone. Thereby, the Bluetooth data give a measure of the proximity between two subjects\footnote{Available at \url{http://realitycommons.media.mit.edu/realitymining.html}.}. This proximity can be interpreted as a link between two subjects. As the time of proximity is also recorded, one can produce a weekly empirical temporal network by grouping the links per week. In this way, a dense network with an average of $9.07$ links per node is obtained.\\
The study of international trade before the $1950$s is hampered by the limitations imposed by the scarcity of data. Thanks to a technique developed in \cite{standaert2015historical}, a reliable coverage of the data on international trade between $1880$ and $2011$ could be accomplished. Note that during the world wars data collection on trade was almost halted. Hence, we exclude these periods from the sample. We construct a temporal international trade network with countries as nodes and establishing links whenever the countries have a significant level of trade integration in a specific year. We treat the international trade data as undirected in order to make a change point analysis with the GHRG method possible.\\
For the Enron e-mail and MIT proximity networks we consider all nodes (including those with no links) in the time windows. For the international trade network, however, we retain the nodes with at least one link throughout the window. In this way a more dense network is obtained, creating improved conditions for change point detection.
\\

\begin{figure}[htb]
	\begin{center}
     		\includegraphics[trim={1.1cm, 0.6cm, 0.8cm, 0.4cm}, clip, width = 0.9\textwidth]{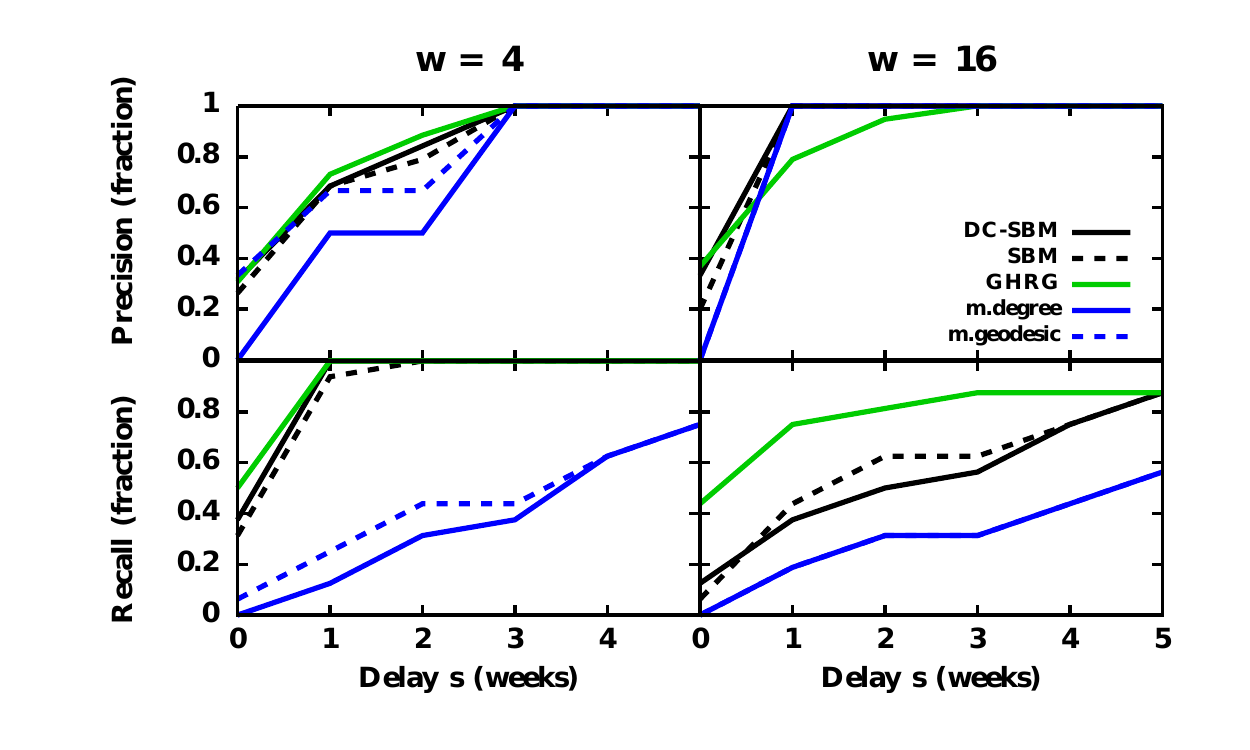}
	 	\caption[Precision and recall. (MIT)]{As in Figure~\ref{enron_precrec} but for the MIT proximity network.
        \label{mit_precrec}}
	\end{center}
\end{figure}
%
\begin{figure}[htb]
	\begin{center}
		\includegraphics[trim={2.7cm, 1.9cm, 3.8cm, 0.8cm}, clip, width=0.65\linewidth]{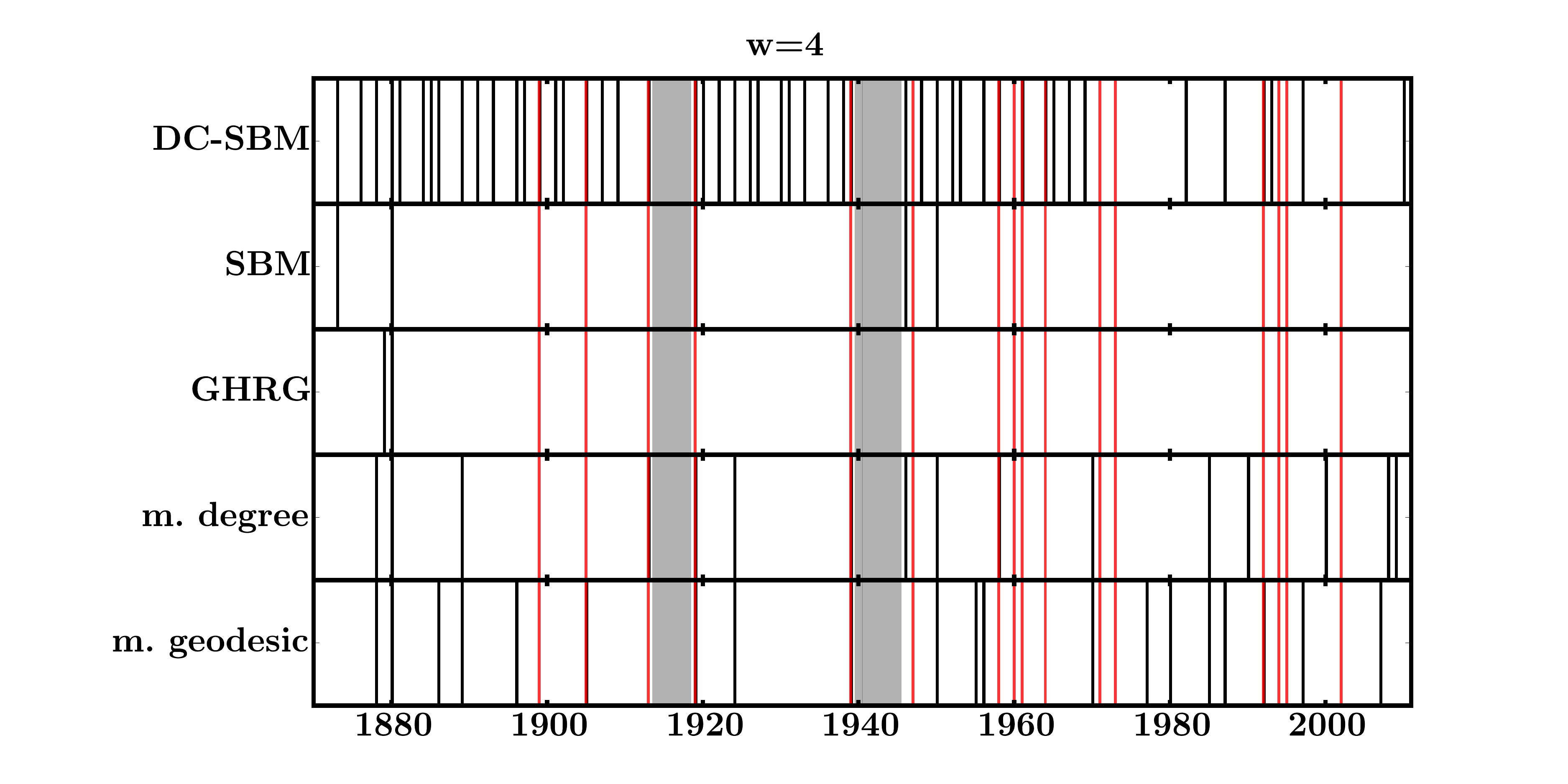}
		\includegraphics[trim={2.7cm, 1.0cm, 3.8cm, 0.3cm}, clip, width=0.65\linewidth]{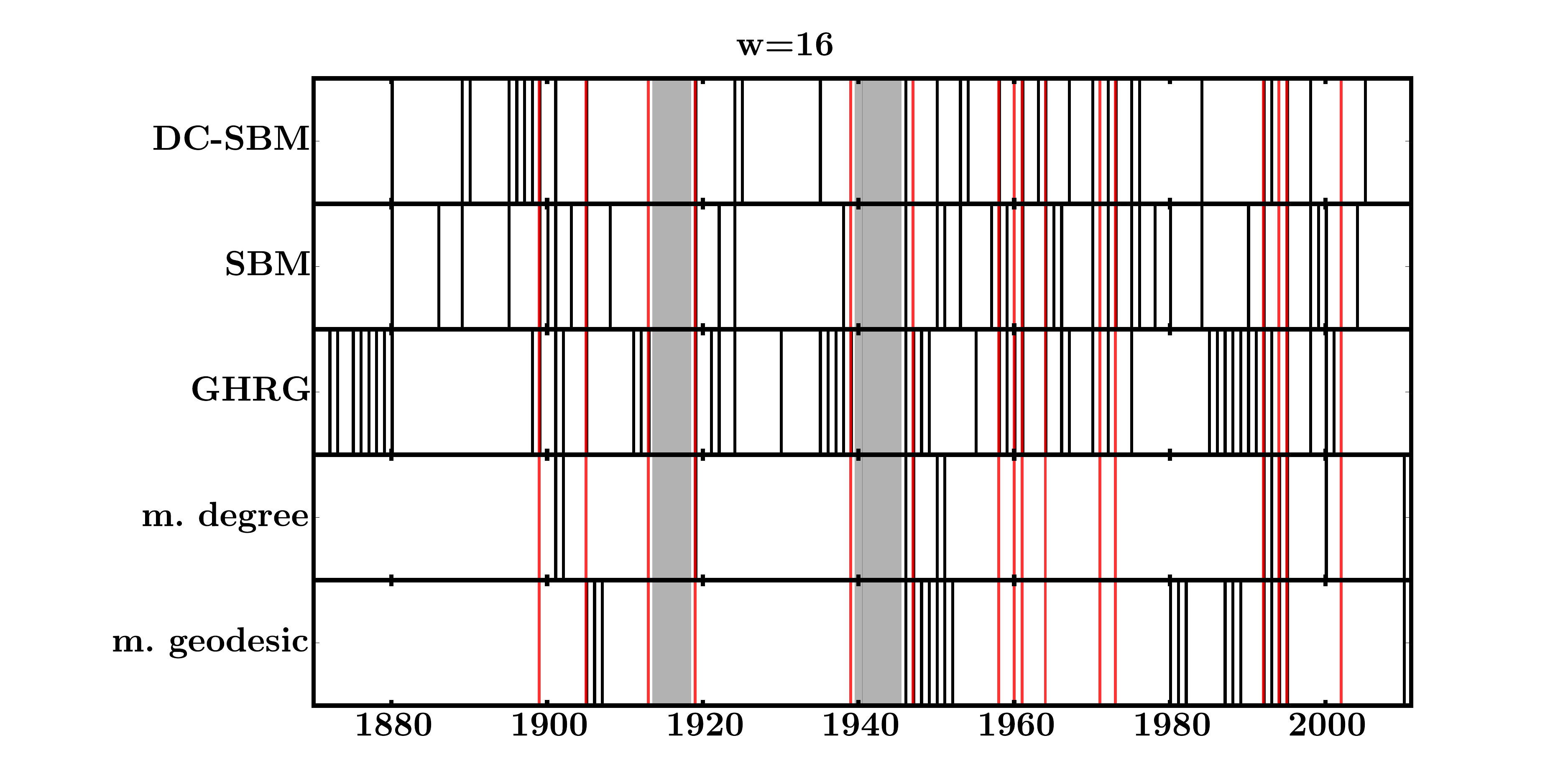}
		\caption[The change points. (Trade)]{As in Figure~\ref{enron_cp} but for the international trade network. The widths of the sliding time windows are expressed in years.
        \label{trade_cp}
        }
	\end{center}
\end{figure}

For each of the three considered temporal networks, there are a number of known dates corresponding with events that are likely to have impacted the network's structure. We treat those dates as if they were the ``empirical'' change points, realizing that they merely mark dates with an enhanced likelihood for changes in the network to occur. The major purpose of the introduction of ``empirical'' change points is to develop a quantitative measure to compare the figure of merit of the different change point detection methodologies. In order to quantify the quality of the various change point detection techniques, we use the ``precision'' and ``recall'' in function of a delay $s$ as it was introduced in \cite{Peel15}
\begin{align}
\operatorname{Precision}(s)=\frac{1}{N_{found}}\sum_{i}{\delta\left(\min_{j}{\left|t^{found}_i-t^{known}_j\right|}\le s\right)}\\
\operatorname{Recall}(s)=\frac{1}{N_{known}}\sum_{j}{\delta\left(\min_{i}{\left|t^{found}_i-t^{known}_j\right|}\le s\right)} \; ,
\label{precrec}
\end{align}
where $N_{found}$ ($N_{known}$) is the total number of detected (``empirical'') change points.  The precision is the fraction of detected change points $t^{found}_i$ that have an ``empirical'' event $t^{known}_i$ within a time range of $s$. The recall is the fraction of ``empirical'' events that have a detected change point within a time range of $s$.
\\

\begin{figure}[htb]
	\begin{center}
	 		\includegraphics[trim={1.1cm, 0.6cm, 0.8cm, 0.4cm}, clip, width = 0.9\textwidth]{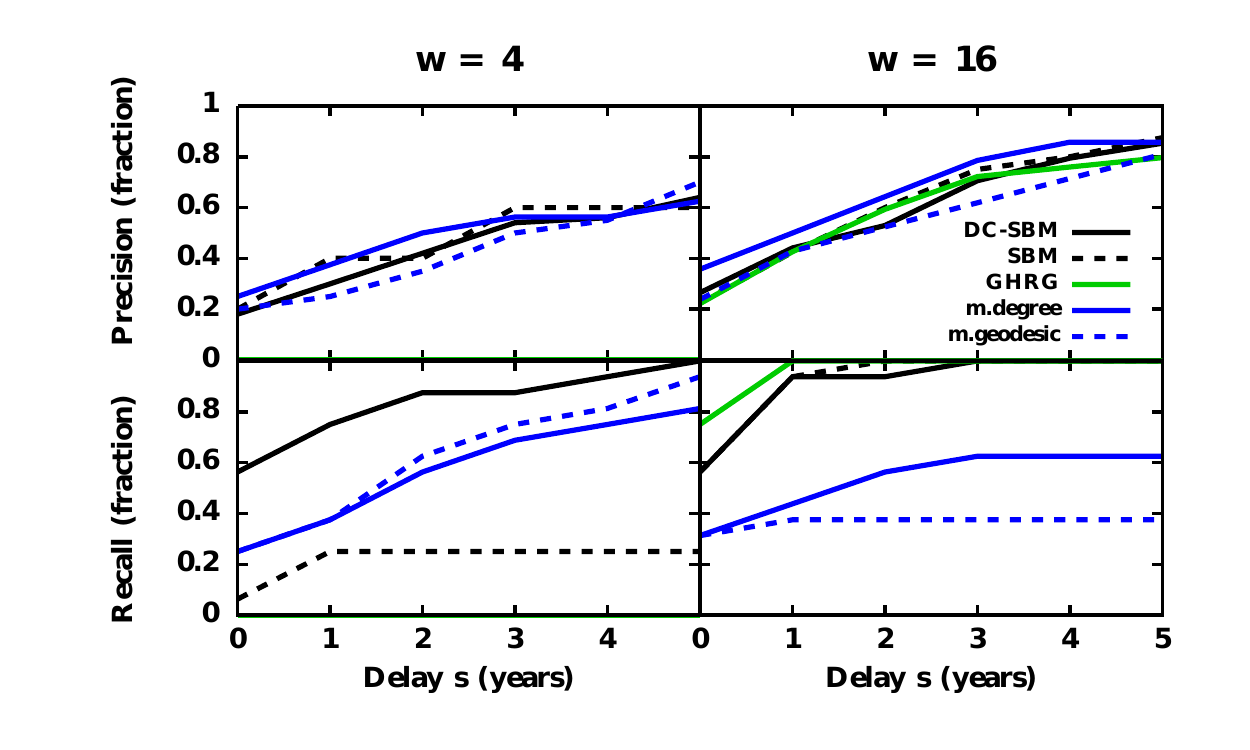}
            \end{center}
		\caption[Precision and recall. (Trade)]{As in Figure~\ref{enron_precrec} but for the international trade network. The widths of the sliding time windows are expressed in years.
        \label{trade_precrec}}
	\end{figure}

Figures \ref{enron_cp}, \ref{mit_cp} and \ref{trade_cp} show the ``empirical'' and the detected change points for the Enron, MIT and trade networks for two different time window widths. The corresponding results for the precision and recall are contained in Figures~\ref{enron_precrec}, \ref{mit_precrec} and \ref{trade_precrec}. In order to get a better feeling of the effect of the width of the sliding time window in the change point searches,  for each temporal network we have been running the algorithms for a ``small" width of 4 ($w=4$) and a ``larger" width of 16 ($w=16$).

When it comes to detecting the ``empirical'' change points, we find that the DC-SBM method is at least equally efficient as the SBM. Furthermore, we observe a strong sensitivity of the detected change points to the value of $w$.
For example, whereas the SBM and DC-SBM predict more change points than the GHRG for  the Enron($w=4$), Enron($w=16$) and MIT($w=4$) combinations, just the opposite is observed for the other three combinations.
This illustrates the sensitivity of the algorithms to the choice made with regard to the value of $w$.
\\

One also faces some situations where the algorithms fail to detect the ``empirical'' change points.
In other situations the algorithms predict a high density of change points, whereas there are no direct empirical indications that point into that direction.
For example, for the international trade network, the combination DC-SBM with $w=4$ leads to many detected change points.
One could argue, however, that 4 years is too small a window for dramatic changes in the international trade network to occur.
For the MIT proximity network, on the other hand, all methods are performing badly for the $w=16$ option.
Here, one could argue that a time window of 4 weeks is a more natural choice to detect changes in the proximity network.
\\

The precision of the various methods for the Enron e-mail network (Figure~\ref{enron_precrec}) is roughly the same.
The GHRG method outperforms the other methods at larger window sizes.
The SBM methods, in particular the DC-SBM version, perform better for the recall.
The simple mean-degree and mean-geodesic methods have a decent precision but lag behind for the recall.
For the precision and recall for the MIT proximity network (Figure~\ref{mit_precrec}), the GHRG method (\cite{Peel15}) displays a slightly better precision, but the SBM methods are slightly better at  recall.
Again, the simple mean-degree and mean-geodesic methods perform well for the precision but are worse for the recall.
For the computed precision of the international trade network (Figure~\ref{trade_precrec}) all methods perform comparably.
For the recall at $w=4$, however, only the DC-SBM method performs better than the local methods.
For a larger window size, both the SBM methods and the GHRG method perform very well.
\\

When comparing our results for the Enron e-mail and the MIT proximity networks with those of \cite{Peel15}, we note some differences, especially for the Enron network.
We see three possible explanations, which may together constitute a plausible explanation.
Firstly, the original datasets were preprocessed in order to turn them into temporal networks.
For the Enron data, a person uses several e-mail aliases, inducing uncertainties in the preprocessing of the data. Secondly, the choice of the time window width is not specified in \cite{Peel15} and, as shown above, the results for the change point candidates depend on that choice.
Thirdly, the detected change points are sensitive to whether only active nodes or all nodes are included in the sliding time window.


\section{Conclusion}
The pioneering work of \cite{Peel15} developed a framework to detect change points in temporal networks based on GHRGs.
In this paper we extend their methodology by adapting it to the use of SBMs as a parametric family of probability distributions for the reconstruction of empirical networks.
We have made a comparative study of the detected change points on three prototypical empirical temporal networks using the GHRG and SBM based methodologies. We have done this for different sizes of the sliding time window and  have also included two more simple change point detection methods in the comparison.
\\

We find that the GHRG method  and SBM methods are  comparably effective in identifying the change points. 
In some sense, the SBM is more versatile in that it can also deal with directed networks for example.
No systematic conclusions could be drawn for the density of the detected change points.
Whereas the SBM models detect more change points than the GHRG for the combinations Enron($w=4$), Enron($w=16$), MIT($w=4$), just the opposite is found for the other three combinations analysed in this work.
This also indicates that the choice of the size of the sliding time window affects the detected change points. When comparing the SBM and DC-SBM methodologies, the DC-SBM version has the tendency to identify a larger amount of change points.
We also find some situations in which the methodologies (even dramatically) over- or under-predict the amount of ``empirical" change points.
Note that the SBM and GHRG are very similar models, as  for an appropriate value of the number of blocks $K$ an SBM equivalent to any GHRG can be constructed. The main difference between the two models being that the GHRG automatically determines the number of blocks at the cost of only being able to recursively partition along the block diagonal of the adjacency matrix. The SBM on the other hand can freely parametrise the full block structure but requires the number of blocks $K$ to be specified. Given the similarity between SBM and GHRG it seems reasonable that they would perform similarly overall, but perform differently for different types of changes. In future work, it may be worth 
partitioning the problem space in more detail so that one can identify  for which types of network changes the various methods perform best.
\\

With respect to the precision and the recall, we conclude that the SBM method produces a better recall than the GHRG method, especially for sparse networks in combination with a ``small" window size.
The precision is only significantly outperformed by the GHRG method  for one of the three studied networks.
In general, the simple mean-degree and mean-geodesic methods do reasonably well for the precision but are outperformed by the sophisticated  GHRG and SBM methods for the recall.
This leads us to conclude that SBMs, and especially the degree-corrected SBM, are a good versatile tool for inference and analysis of complex networks.
The inference of change points in temporal networks, however, is subject to some uncertainties which are connected with the adopted method and the widths of the considered sliding time windows.
Methodologies based on parametric families for reconstructing the empirical networks, however, outperform the more simple methodologies.  
\\

An implementation of the proposed algorithm is available at \url{https://github.ugent.be/pages/sidridde/sbm_cpd}.
The independence between the runs in the different time windows makes parallelisation easily attainable.
In each time window, the sparse version of the belief propagation algorithm leads to a computational complexity of $\mathcal{O}((MN+N^2)K^2)$, and a memory complexity of the order $\mathcal{O}(M)$.

\bibliographystyle{iopart-num}
\bibliography{bibliography}

\end{document}